\documentclass[aps,prstper,preprint,groupedaddress]{revtex4-1}

\pdfoutput=1
\usepackage{amsmath} 
\usepackage{amsfonts} 
\usepackage{graphicx}
\usepackage{tikz}
\usepackage{xcolor}
\usepackage{hyperref}

\begin{document}

\title{Interactions between teaching assistants and students boost engagement in physics labs}

\author{Jared B. Stang}
\email{jstang@phas.ubc.ca}
\affiliation{Department of Physics \& Astronomy, University of British Columbia, 6224 Agricultural Road, Vancouver, British Columbia V6T 1Z1 Canada}

\author{Ido Roll}
\email{ido.roll@ubc.ca}
\affiliation{Centre for Teaching, Learning, \& Technology, University of British Columbia, 214-1961 East Mall, Vancouver, British Columbia V6T 1Z1 Canada}

\date{\today}

\begin{abstract}
Through in-class observations of teaching assistants (TAs) and students in the lab sections of a large introductory physics course, we study which TA behaviors can be used to predict student engagement and, in turn, how this engagement relates to learning. For the TAs, we record data to determine how they adhere to and deliver the lesson plan and how they interact with students during the lab. For the students, we use observations to record the level of student engagement and pre- and post-tests of lab skills to measure learning. We find that the frequency of TA--student interactions, especially those initiated by the TAs, is a positive and significant predictor of student engagement. Interestingly, the length of interactions is not significantly correlated with student engagement. In addition, we find that student engagement was a better predictor of post-test performance than pre-test scores. These results shed light on the manner in which students learn how to conduct inquiry and suggest that, by proactively engaging students, TAs may have a positive effect on student engagement, and therefore learning, in the lab.
\end{abstract}


\maketitle



\section{Introduction}
\label{intro}



\subsection{Background}
In science courses at the university level, teaching assistants (TAs) take on a variety of roles and are integral to the successful delivery of many courses.\cite{Seymour:2005} These roles range from marking assignments and proctoring exams to facilitating laboratories and leading tutorials. In large-scale courses, as are typical for first-year undergraduates at large universities, TAs are especially important for learning,\cite{Seymour:2005} as in many cases they serve as the sole instructors in the smaller lab or tutorial sections. 

\begin{figure}[h!]
\centering
\includegraphics[width=0.6\linewidth]{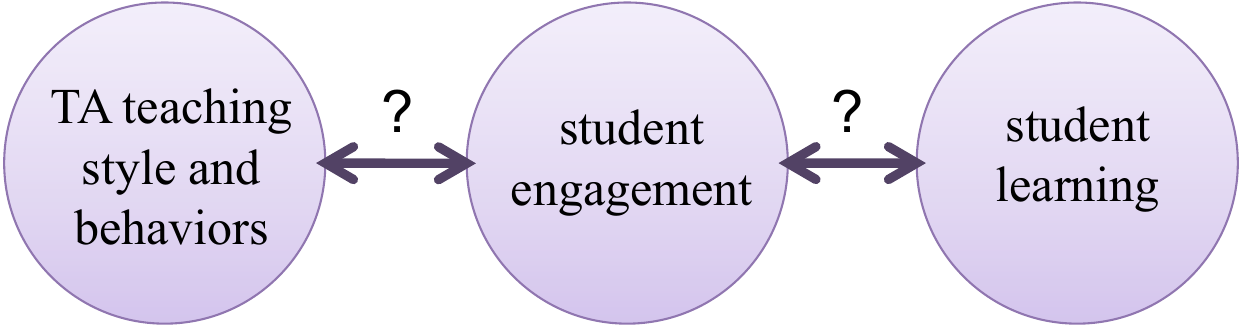} 
\caption{A simplified model of the TA--student relationship. Each bubble is hypothesized to influence the bubbles immediately next to it. In this paper, we evaluate the correlational links between `TA teaching style and behaviors' and `student engagement' and between `student engagement' and `student learning'. The results are summarized in Figure~\ref{TA_student2}.}
\label{TA_student}
\end{figure}

Given the large amount of responsibility TAs have in these courses, the question becomes, what effect do the TAs have on student learning? In this study, we focus on two questions of interest: How do TA behaviors in the lab affect student engagement and, in turn, what is the relationship between this engagement and learning in the physics lab? (See Figure~\ref{TA_student}.) As explained further below, our focus on student engagement stems from the fact that engagement has been found to be a significant predictor of learning.\cite{kuh2008unmasking,pace1990undergraduates} Studying these relationships will contribute to an understanding of the wider question of how TAs affect students and student learning in the classroom and may provide evidence for the effectiveness of particular instructional strategies.


Clearly, there are many other factors, not included in the model of Figure~\ref{TA_student}, that influence student engagement and learning. For example, it may be that a student is very motivated in a particular course due to a personal interest in the material and not because of the actions of the TAs. Also, TA behaviors may affect learning directly, not through engagement. For example, an enthused or clear TA may be more effective. Despite these limitations, this model conveys a simplified representation of the relationship between TAs, students, and learning that is the focus of our current investigation.

We undertake this study in the lab sections of a transformed large-scale first year algebra based physics course. We begin by reviewing existing literature on TA behaviors and student engagement and providing further motivation for this work. We then describe the context for our investigation and outline the experimental design. Following the results, we discuss the implications for TA training in first-year large enrolment labs. 

\subsection{Literature review and motivation}
Research has clearly shown that variations in TAs' beliefs and attitudes influence their behavior as an instructor.\cite{Goertzen:2009,Singer:1996,Trigwell:1996,Spike:2010,Spike:2010a,Goertzen:2010,Speer:2005,Speer:2008} Subsequently, recent studies have examined the behaviors of TAs, reporting for example on who initiates TA--student interactions\cite{Scherr:2006} and the variations in the types, frequencies, and lengths of TA--student interactions that occur.\cite{West:2012,Patitsas:2012,debeck:2012} Despite this focus on the actions of TAs and instructors, the literature has not yet made connections between these behaviors and the student learning process and outcomes. In this study, we seek to understand the potentially important effect of these different TA modalities (such as interaction lengths and whether or not the TAs initiate interactions) on students in the physics lab, thereby addressing this open question.

Student engagement in college classrooms may be conceptualized as ``the time and effort students devote to activities that are empirically linked to desired outcomes of college."\cite{kuh2009student} Engagement of this general type has been shown to be positively linked to grades and persistence in undergraduate students.\cite{kuh2008unmasking} The above definition of engagement includes many dimensions\cite{chickering1987seven} and applies across the spectrum of undergraduate experiences. To define engagement in the context of the physics lab, we focus on the dimension of being on task. Thus, we define an engaged student as one that is visibly focussed on and occupied with the lab activity. This type of engagement has been shown to positively correlate with learning and grades in the lecture,\cite{duncan2012digital,sana2013laptop} in the use of intelligent tutoring software,\cite{baker2004off} and in general.\cite{pace1990undergraduates} 
Student engagement, as defined here, is a variable which only exists during class time. As our unit of observation for both TA behaviors and student engagement is then the single lab period (see section \ref{method}), this allows us to evaluate the possible impact of TAs on students at the same level. The importance of engagement for learning and the practical  consideration of evaluating a student outcome in the single lab period motivate our focus on student engagement.


The second question in our study investigates the relationship between engagement, as defined above, and learning. The lab in which this study is undertaken is a hands-on inquiry-based lab. Then, in order to learn, students are expected to engage with the lab content, as there are no other opportunities for them to participate in the learning activities (for example, through lecture notes at home). We therefore hypothesize that in this context engagement will be correlated with learning. Results connecting engagement with learning in this specific situation will help to understand the effect of TA behaviors on student learning and contribute to the existing literature, reviewed above, on the general positive relationship between student engagement and learning.

%
%


\section{Method}
\label{method}
\subsection{Design}

This study took place in the lab sections of a large-scale first year physics course. Within the inquiry-based lab, in which the TAs are the sole instructors, we look in particular at two aspects of TA teaching style and behaviors: {\it how the TAs adhere to and deliver the lesson plan}\footnote{For each lab, the TAs are provided a `TA version' of the lab activity, with timing notes and comments about facilitation. See appendix \ref{TA_copy} for an excerpt from the TA copy of the lab worksheet.} and {\it how the TAs interact with students during the work session}.
The specific TA behaviors we record (in italics) are as follows: 
\begin{itemize}
\item How the TAs adhere to and deliver the lesson plan.
\begin{enumerate}
\item {\it Behaviors outside of the standard TA script}.
\end{enumerate}
\item How the TAs interact with students during the work session.
\begin{enumerate}
\setcounter{enumi}{1}
\item {\it Number of interactions}.
\item {\it Who initiates the interactions}.
\item {\it Length of interactions}.
\end{enumerate}
\end{itemize}
Our main strategy for data collection was to perform observations of the lab. From an unobtrusive vantage point at the back of the lab room, observers were able to record data on the TA behaviors listed above. 

This study further evaluates two student factors: {\it engagement} and {\it learning}. We define an engaged student as one that visibly has their attention focussed on the lab activity. As behavioral measures are often more reliable than self-reports,\cite{macleod:1996,Roll:2008} we measure engagement through observations. By periodically circulating the room, observers measured snapshots of student engagement in the lab. As discussed in more detail in section \ref{data_col}, strategies to minimize the disturbance of the observations on the classroom were utilized. The observations were completed during one lab session in a typical week in the middle of the term. Finally, learning was measured by giving students a lab exam on the first and last weeks of the term. 

\subsection{Participants and description of lab}

We undertake this study in Physics 100, a first-year introductory physics course at the University of British Columbia (UBC), a large research-intensive university. 713 students were spread between 17 lab sections. The students in this course are primarily in the life sciences and, for most of them, this will be their only physics course. In addition, the majority of the students are first-year first-term university students.

The lab portion of Physics 100, which consists of weekly sessions, each 80 minutes in length, has been extensively revamped in recent years to utilize an inquiry-based approach. The learning goals for the lab are aimed at the development of a general set of scientific and data analysis tools rather than being focused on content knowledge. A typical lab begins with an introduction and a set of clicker questions, followed by an extended period of students working in pairs, before ending with a summary discussion and another set of clicker questions. The average number of students in each lab was 39 while the largest and smallest sections had 46 and 25 students in them. 

During the observation week, the main task of the lab was to collect and extrapolate a data set in order to make a prediction for a future experiment. By measuring the time it took a coffee filter to fall to the ground from heights below 1 metre, students had to predict the time it would take for the coffee filter to fall from a height of 2 metres. To complete this inquiry-based activity, students were required to design and carry out an experiment that would allow them to predict the falling time from the desired height, for example by performing a linear fit to data for various heights below 1 metre. Once each pair of students had made a prediction, one TA would perform (or get students to perform) the experiment at the height of 2 metres while the other would collect the predictions from each group. We define the `student working period' in the lab as the time during which students are working to make their predictions, beginning after the introduction and ending once the TAs begin performing the experiment at a height of 2 metres or soliciting student predictions. The lab ended with a discussion comparing the predictions to the measured results.

In Physics 100, the TAs are the instructors in the lab, having full control of the section. We exclude one of the lab sections from our analysis, because in the observation week there was a replacement TA, as lab norms that may have been established with the regular TA may have been different than the results of our observations. A group of 10 TAs, consisting of 9 males and 1 female, facilitated the 16 remaining lab sections in pairs. Each pair of TAs were assigned to 2 consecutive lab sections, and each TA was assigned to 2 or 4 sections. A TA assigned to 4 sections then taught two different instances of consecutive labs, with a different TA partner for each instance. The TAs attended weekly meetings in which the upcoming lab was reviewed.

6 of the 10 TAs were first-year graduate students at UBC and had no prior teaching experience, while 2 TAs had more than one year of TA experience at UBC. The TAs ranged in age from 22 to 28, with a median age of 24. Upon beginning graduate school at UBC, each of the first-year TAs underwent an 8 hour TA Professional Development Workshop in addition to a 3 hour Physics 100 specific workshop.\cite{holmes:2013}

\subsection{Data collection and analysis}

\subsubsection{Data collection}
\label{data_col}

To observe the TAs, we developed the \textit{TA observation form} (appendix \ref{TA_form}) to record a timeline of the TAs' activities during the lab. On the form, we identified three main TA behaviors: 
\begin{enumerate}
\item `Talking to class': The TA is addressing the entire class (for example, leading a classroom discussion).
\item `Inactive': The TA is not available to students (for example, the TA may be out of the room or talking to the other TA).
\item `Active': The TA is either helping students or is available to help students.
\end{enumerate}
Under the `Active' category, the number, length and initiator of interactions with students was recorded. In addition, our form allowed us to record other section data, including the progression of the lab (through the introduction, clicker questions, etc.). A category titled `extras' was used to record any TA behavior that was outside of the main script of the lab. (See appendix \ref{extras} for a list of extra behaviors recorded.) A completed TA observation form allows us to see what the TA was doing at each moment during the observed lab.

To measure student engagement, we adapted the Baker-Rodrigo Observation Method Protocol,\cite{ocumpaugh2012baker,fieldnotes_engagement} developing the \textit{On/off task form} (appendix \ref{oo_form}). This form consisted of a spatial map of the lab with empty squares to represent student positions at each lab bench. To fill out the form, the observer looked at the lab bench. If a student was on-task, as judged by the observer at a glance, a check-mark was placed in the corresponding square on the form while if a student was off-task, an `x' was placed in the square. Some clues observers used to assign the binary engagement value to each student include whether or not the student was involved in an off-task behavior and whether students were attending to their partner and their task through eye contact, body position and verbal cues. For example, a student that appeared to be writing on their lab worksheet was judged as on-task and a student that was looking at their cell phone was marked as off-task. The On/off task form was completed at intervals of ten minutes during the lab, giving snapshots of the engagement level of the class. The fractional engagement for a snapshot is defined as the number of on-task students (number of check-marks) divided by the total number of students present in the section. Both the TA observation form and the On/off task form underwent iterative design.

Four observers observed 7-9 sections each. Two observers observed each lab section, with each observer documenting the actions of one TA. A set of practice observations were undertaken prior to the observation week and used to confirm inter-rater reliability. During this 15 minute practice observation, all four observers recorded 90-93\% agreement on the number of interactions ($12.5\pm1.3$ recorded interactions) and the number of short interactions (less than one minute long) ($11.0\pm0.8$). Far fewer long interactions ($1.5\pm0.6$) and `extra' off-script TA behaviors ($2.0\pm0.8$) were recorded and as such these categories did not show as good of an agreement. One snapshot of the student engagement level of the class was taken, giving a fractional student engagement of $0.87\pm0.06$ between the observers, showing agreement of 93\%. All disagreements between the observers were on two benches which appeared off-task by two of the observers but on-task by two other observers. Therefore, there is agreement within the observers on which benches were off-task and it appears that most of the variability in fractional engagement is due to the differences in the precise time at which the students were observed. We expect that such effects will be mitigated by recording multiple snapshots.

To record data on the TA observation form, the observers watched the TA from an unobtrusive vantage point at the back of the classroom. In order to not interfere with the lab, observers made a conscious effort to not talk or interact with the TAs or students in the lab. Further, the TAs were told only that the observers would be observing the classroom. They were not told that the observers would be recording specific data about the TA, nor were they told that the observers were not. We note also that these labs were often observed by visitors who were unfamiliar to most TAs and students. These design choices were made so that the effect of observers is not different from other observers who frequently attend the lab, and thus should not affect the natural behaviors of students or TAs. Due to the observation style, any data about the style, content and quality of the TA--student interactions is outside the scope of this study. To complete the On/off task form required the observer to move from the back of the lab room, in order to properly observe the engagement of the students. Typically, the observer would take 1-2 minutes to walk around the classroom and fill out the form, again consciously avoiding interactions with students and TAs. To avoid an effect in which students in the lab might be motivated to appear to be working if they feel like they are being observed, the observers made an effort to observe the students discreetly. One strategy in this regard, previously used by Baker {\it et al},\cite{ocumpaugh2012baker,Baker:2008} was to stand near one lab bench while observing a different bench.

To evaluate the relationship between engagement, interactions and learning, we evaluated student learning using pre- and post-lab exams. The lab exams were given during the first and last week of the lab. The exams were adapted from the Concise Data Processing Assessment\cite{Day:2011} and Lawson's Classroom Test of Scientific Reasoning.\cite{Lawson:1978} The tests were given without prior notice. All items were multiple choice, as shown in appendix \ref{lab_test}.

\subsubsection{Analysis}

We compiled the TA data to give results on a per-lab basis, as student engagement is expected to be a product of the behaviors of both TAs in a section. For the number of TA--student interactions and the student engagement measurements, we use data from the `student working period' only. This is the only period during the lab in which these measurements are defined and relevant. In addition, from the perspective of the TAs, this `student working period' is rather free-form. Within the main goal of supporting the students' activities, differences in TA facilitation style are expected to manifest themselves as different TA behaviors during this time. Since the time each lab section spends in this `student working period' varies, in order to compare across sections the number of TA--student interactions are normalized by the length of the section's `student working period', giving a frequency of interactions. The calculation of this frequency takes into account moments in which observers were not observing due to a variety of reasons. (Usually, this was due to observers completing the On/off task form.) 

The length of interactions and the frequency of interactions are interdependent, as the longer the interactions are, the fewer interactions one has time to do. In order to disentangle the effect of the length of interactions from the number of interactions, we look at the ratio of short to long interactions, which gives a unit-less measure that does not depend on the frequency of interactions. Through the 16 sections, 564 TA--student interactions were recorded. Of these, 391 were less than a minute. Therefore, we have taken interactions lasting greater than or equal to a minute to be `long', with interactions lasting less than a minute being `short'.

We recorded the number of behaviors outside of the standard TA script through the entire lab. 

In these labs, it is customary to move on only when all students have completed the task, so that there is typically some down time for quick students. It is important that we measure engagement during the time in which students choose to be on task, before students have finished their lab activity. We restrict the engagement results to the relevant part of the `student working period' by taking only the first three student engagement snapshots recorded. These three snapshots span a time of $19.6\pm3.3$ minutes. 

Finally, we have the student learning data. As we combine TA data by section, we average the lab test performance across all students in each section. Overall, students improved significantly on test items that were shared by the pre-
and post-tests. Pre-test: $66\%\pm 5\%$; post-test: $76\%\pm4\%$. $t(15)=10.1$, $p<0.0001$.

\subsection{Limitations}
\label{limitations}

One of the main limitations of this study is that the observational data is restricted to a single lab. This is especially challenging for examining the link between engagement and learning, as we correlate measures from a single snapshot (student engagement) with outcomes of the overall lab course (student learning). Thus, our study does not take into account the week-to-week variations that occur in many important factors, such as TA behaviors, lab style, student attitudes, etcetera. In the context of this study, these variations would likely introduce random noise and weaken possible relationships. Finding a significant correlation between engagement and learning in this highly noisy environment, as we do below (section~\ref{corr_learn}), thus provides evidence that such a relationship does exist. It would be very interesting to study classroom engagement over the course of a semester to provide more information about the possible connection between our single lab observation and semester-long trends. Such study would help to confirm or refute the results presented here.

Another important limitation is our inability to access more fine-grained information about the style and content (and quality) of TA--student interactions. It would be expected that the quality of interactions plays a large role in determining the student response to the TAs. In addition, evaluating the content of the interactions would provide important data about which types of interactions are useful for student learning. For example, does a TA who is sympathetic to students' thinking provide better support for learning?\cite{Scherr:2011} To collect this data would either require the observers to be more intrusive, which would complicate the analysis, or necessitate the use of more advanced technology (i.e video and audio recording), both of which are outside the scope of our study.

A final concern in our design is the possible presence of an observer effect in the student engagement data, as students may tend to make sure they appear to be working if they feel that someone is watching them. The average engagement level through the three engagement snapshots used in our results is very high, at 88\%, and the range is from 67\% to 99\%. Any observer effect would push engagement numbers higher, tending to minimize the differences in the data and washing out the effect. Thus, we expect that our method of collecting engagement data is sufficient and that any correlation between the number of TA--student interactions and engagement will be robust.

\section{Results}
\subsection{Descriptives of TA teaching style and engagement results}

\begin{table*}[h!]
\centering
\caption{The time spent in each stage of the lab for all 16 observed sections. The `student working period', during which students were working in pairs to complete the activity, was the longest stage of the lab.}
\begin{ruledtabular}
\begin{tabular}{l c c}
Lab stage & Mean (m) & $\sigma$ (m) \\
\hline
Discussion of previous homework & 7.9 & 3.1  \\
Introduction and first clicker questions & 11.3 & 2.7  \\
`Student working period': Students working to make predictions & 32.0 & 6.3  \\
TAs perform experiment, class discussion and final clicker questions & 26.3 & 6.6
\end{tabular}
\end{ruledtabular}
\label{breakdown}
\end{table*}

We observed a high amount of variability between the 16 lab sections, as summarized in Tables \ref{breakdown} and \ref{dat}. This wide variation in TA style is evident in the recorded number of behaviors outside the standard TA script ($9.4\pm4.6$). The number of interactions per minute across the sections was $1.31\pm0.42$. 365 (or 65\%) of the 564 interactions were initiated by the TAs,\footnote{The proportion of TA-initiated interactions we recorded agrees with the proportion reported in Ref.~\onlinecite{Scherr:2006}.} while the rest (except for 23 interactions with no marked initiator) were initiated by students. Lastly, the ratio of short to long interactions was $2.8\pm1.3$; the majority of interactions were less than 1 minute long.

The student engagement results show a very high fractional engagement, at $0.88\pm0.08$, through the first three engagement snapshots.

\begin{table*}[h!]
\centering
\caption{TA behavior data descriptives.}
\begin{ruledtabular}
\begin{tabular}{l c c c c}
TA behavior & Mean & Max & Min & $\sigma$ \\
\hline
(\# of) behaviors outside of the standard TA script & 9.4 & 22 & 4 & 4.6 \\
Number of interactions per minute & 1.31 & 2.33 & 0.51 & 0.42 \\
Fraction of interactions initiated by the TA & 0.63 & 0.92 & 0.39 & 0.16 \\
Ratio of short ($<$ 1 minute) to long ($\geq1$ minute) interactions & 2.8 & 5.25 & 0.46 & 1.3
\end{tabular}
\end{ruledtabular}
\label{dat}
\end{table*}

\subsection{Correlations}
\subsubsection{Correlation of TA behaviors with student engagement}

\begin{table*}[h!]
\centering
\caption{Correlation of TA behaviors with student engagement. At an $\alpha$ level of $0.05$, the frequency of interactions and the frequency of TA-initiated interactions give the only statistically significant correlations.}
\begin{ruledtabular}
\begin{tabular}{l c c}
TA behavior & $r(14)$ & $p$ \\
\hline
(\# of) behaviors outside of the standard TA script & -0.11 & 0.66\\
Number of interactions per minute & 0.52 & 0.03 \\
- Frequency of TA-initiated interactions & 0.49 & 0.04 \\
- Frequency of student-initiated interactions & 0.052 & 0.84 \\
Ratio of short ($<$ 1 minute) to long ($\geq1$ minute) interactions & -0.10 & 0.70
\end{tabular}
\end{ruledtabular}
\label{corrs}
\end{table*}

\begin{figure}[h!]
\centering
\includegraphics[width=0.6\linewidth]{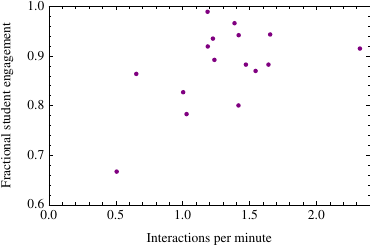}
\caption{Fractional student engagement versus number of TA--student interactions per minute during the `student working period' in the lab. The frequency of interactions is positively correlated with engagement: $r(14)=0.52$.}
\label{int_eng}
\end{figure}

We correlated student engagement with each of the TA behavior variables, using an $\alpha$ level of $0.05$. These results are shown in Table~\ref{corrs}. As shown in Figure~\ref{int_eng}, the frequency of interactions between TAs and students was a positive and significant predictor of student engagement: $r(14)=0.52$, $p=0.03$. $r^2=0.27$, suggesting the 27\% of the variability in student engagement per section is explained by the frequency of their interactions with TAs. In addition, the frequency of interactions initiated by TAs is significantly correlated with engagement, $r(14) = 0.49$, $p = 0.04$. The frequency of interactions that were initiated by students is not correlated with engagement: $r(14) = 0.052$. The other variables were not significantly correlated with student engagement. 

\subsubsection{Correlation of student engagement with learning}
\label{corr_learn}

To evaluate the relationship between student engagement and learning, we calculated the partial correlation with the post-test, controlling for performance on the pre-test. 
 The measure of engagement is significantly correlated with learning: partial-$r(13) = 0.56$, $p = 0.03$, $r^2 = 0.31$. Thus, $31\%$ of studentsÕ performance in the lab can be explained by their engagement (as measured in a single session). Performance on the pre-test by itself is not significantly correlated with post-test scores: $r(14) = 0.32$, $p = 0.23$.

\section{Discussion}
\subsection{Discussion of results} 

\begin{figure}[h!]
\centering
\includegraphics[width=0.6\linewidth]{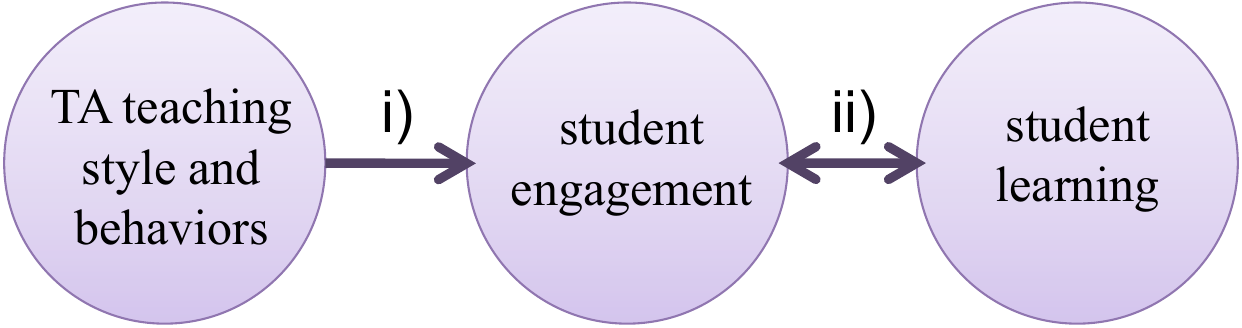} 
\caption{The main results of our study. We found significant positive correlations (denoted by a solid line) between: i) the frequency of TA--student interactions (a product of `TA teaching style and behaviors') and student engagement ($r(14)=0.52$); ii) student engagement and student learning (partial-$r(13) = 0.56$, controlling for pre-test). The correlation between the frequency of TA--student interactions and engagement is explained solely by TA-initiated interactions, implying that it is the TAs that are driving student engagement (indicated by a unidirectional arrow).}
\label{TA_student2}
\end{figure}

Of the TA behaviors described in section \ref{method} and monitored in our study, only the number of interactions per minute has a significant correlation with student engagement. Interestingly, the correlation between interactions and engagement is explained solely by TA-initiated interactions. The rate of student-initiated interactions had no relationship with engagement, while a high rate of TA-initiated interactions is strongly associated with increased engagement. Finally, we found a significant correlation between engagement and learning in this inquiry-based lab. Our results are summarized in Figure~\ref{TA_student2}.

While our observations are correlational and not causal, suggesting potential explanations for these correlations is of interest. Interactions may positively contribute to motivation and engagement by satisfying students' needs and enabling their success. Also, interactions may reduce disengagement due to a policing effect. The fact that interactions were related to engagement only when initiated by TAs suggests that the relationship is not due to student factors such as intrinsic motivations. Instead, it seems that TA behaviors are the ones that increase engagement, whether perceived as a carrot (due to the given assistance) or a stick (due to their policing effect).

The dependency of the results presented above on the inquiry nature of the lab is of interest. We hypothesize that the correlation between interactions and engagement is not specific to an inquiry-based approach. Our two main inferences, that interactions allow students' needs to be met and that the policing effect of interactions reduces disengagement, are expected to also be valid in a traditional style lab. Compared to an inquiry-based lab, the needs of the students would likely be different in a traditional lab. However, interactions would still allow the TAs to support the students as they work through the lab. Therefore, we speculate that the correlation between interactions and engagement may be generalizable across settings. In contrast, the correlation between engagement and learning may be specific to the inquiry-based lab. In such a lab, students are responsible for generating their own knowledge. If designed properly, there is no avenue for students to learn in the lab without thinking about the process and reflecting on the results, as may be possible in a traditional lab.

It is interesting that the length of interactions does not correlate with the engagement of the students. Although it might be expected that longer interactions would be more positive, as they may be able to better satisfy the students' needs and help the students to feel invested in, our results provide no evidence to support this claim. This suggests that a brief stop by the TA is as effective as an in-depth interaction in keeping students engaged in the lab. A possible explanation is that by initiating a short interaction, the TA opens the door for deep questions from the students, in a sort of `ventilation effect'. In this scenario the length of the interaction is not an outcome of TA style, but rather, an adaptive behavior based on students' needs. Then, the length should not affect engagement, as indeed suggested by the data. However, further study is necessary to discern the true effect, if any, of the length of interactions.

The number of off-script TA events does not correlate with student engagement in the lab, even though one might expect that this sort of adaptive instruction might help students to have the tools they need to work through the lab.\cite{chi:2009} It could be that off-script behaviors might not be personalized enough to motivate the general student. However, it is more likely that our category was too coarse-grained and that only certain types of adaptive instruction have an effect on student engagement. A more detailed study, with the ability to distinguish different types of off-script behaviors, is necessary to establish the specific adaptive behaviors that benefit students.


In addition to the hypotheses tested above, our observations show interesting general results about how TAs spend their time in the class. We observed a large variability in the number of interaction and the number and type of off-script behaviors. In addition to these, there was a substantial variation in the progression of each lab section. It is clear that there are meaningful differences in the facilitation style of different TAs and that these differences are an important factor in determining student engagement and learning. 

\subsection{Conclusions and implications}

The correlation of the frequency of TA-initiated interactions with student engagement is an important first step in studying the effect of TAs on student outcomes in a lab course. The results presented here establish a direct relationship between TA behaviors (initiating interactions) and the student response (engagement), thereby demonstrating that variations in TA facilitation styles do have an effect on students and identifying a particular example of this effect. Furthermore, the positive correlation between student engagement and learning in this lab shows that this effect is (indirectly) related to student learning. 

In addition to better understanding the effect of TA teaching style on student engagement and learning, results from this study inform and support TA professional development efforts in order to encourage more productive tutoring styles. The positive correlation between TA-initiated interactions and student engagement (and, in turn, the positive correlation between student engagement and learning) suggests that TAs should proactively engage with many students. Since the length of interactions was not found to affect the engagement of students in this course, it is possible that just saying `Hello, is everything okay?' may be as useful as a lengthy discussion in keeping students on task. In our opinion, a more likely explanation is that such short interactions are useful in keeping an eye on student needs and enabling a `ventilation effect', giving students an opportunity to access the TAs and opening the door for student questions that turn these brief exchanges into long and meaningful interactions.

\begin{appendix}

\newpage
\section{TA copy of lab}
\label{TA_copy}

In this section, we provide the first three pages of the TA version of the lab worksheet in the observation week. The student version excludes the `TA guidelines' column on the right of the page.

\begin{figure*}[h!]
\centering
\fbox{\includegraphics[scale=0.65]{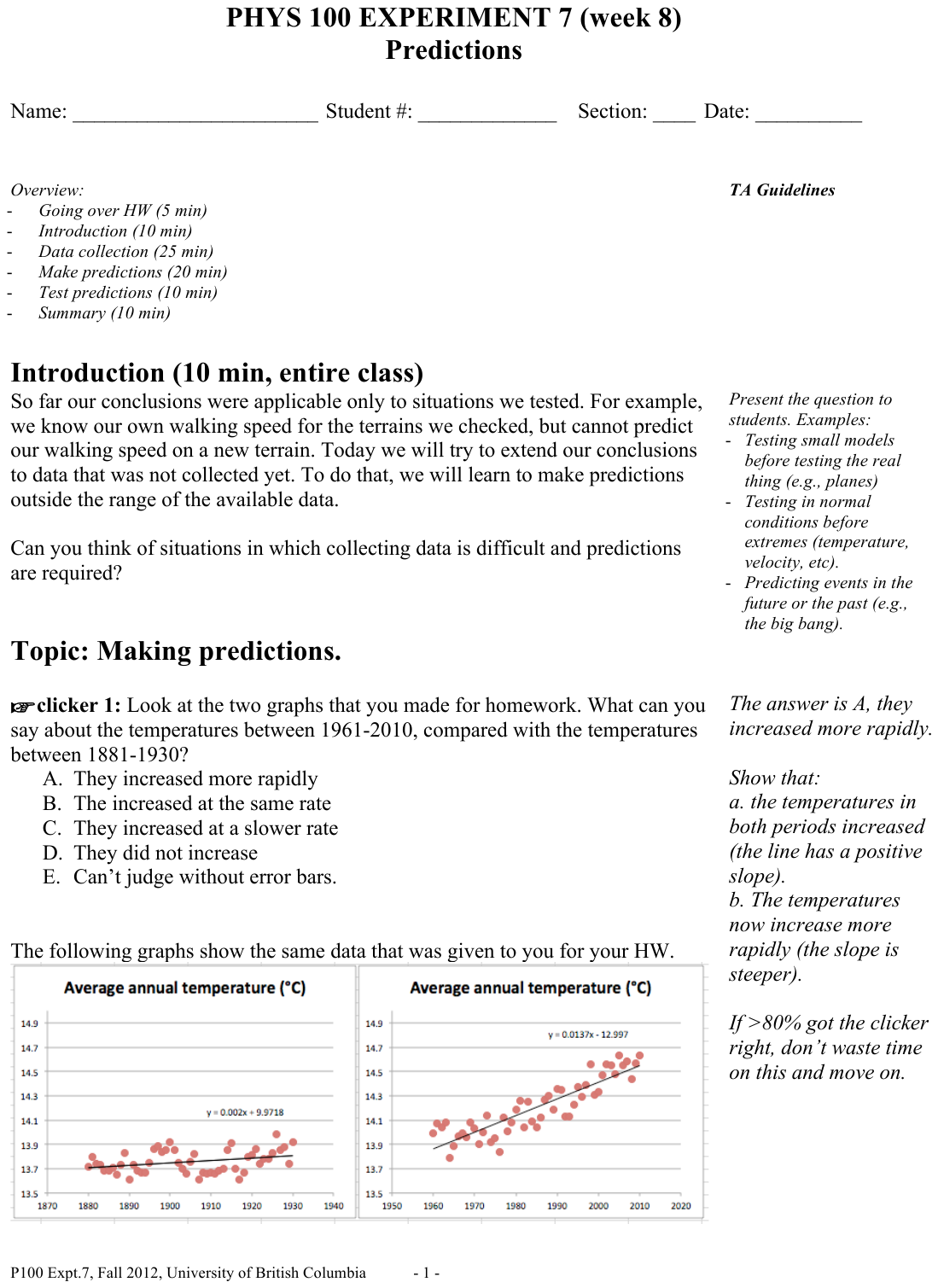}}
\caption{}
\label{}
\end{figure*}

\newpage
\begin{figure*}[h!]
\centering
\fbox{\includegraphics[scale=0.65]{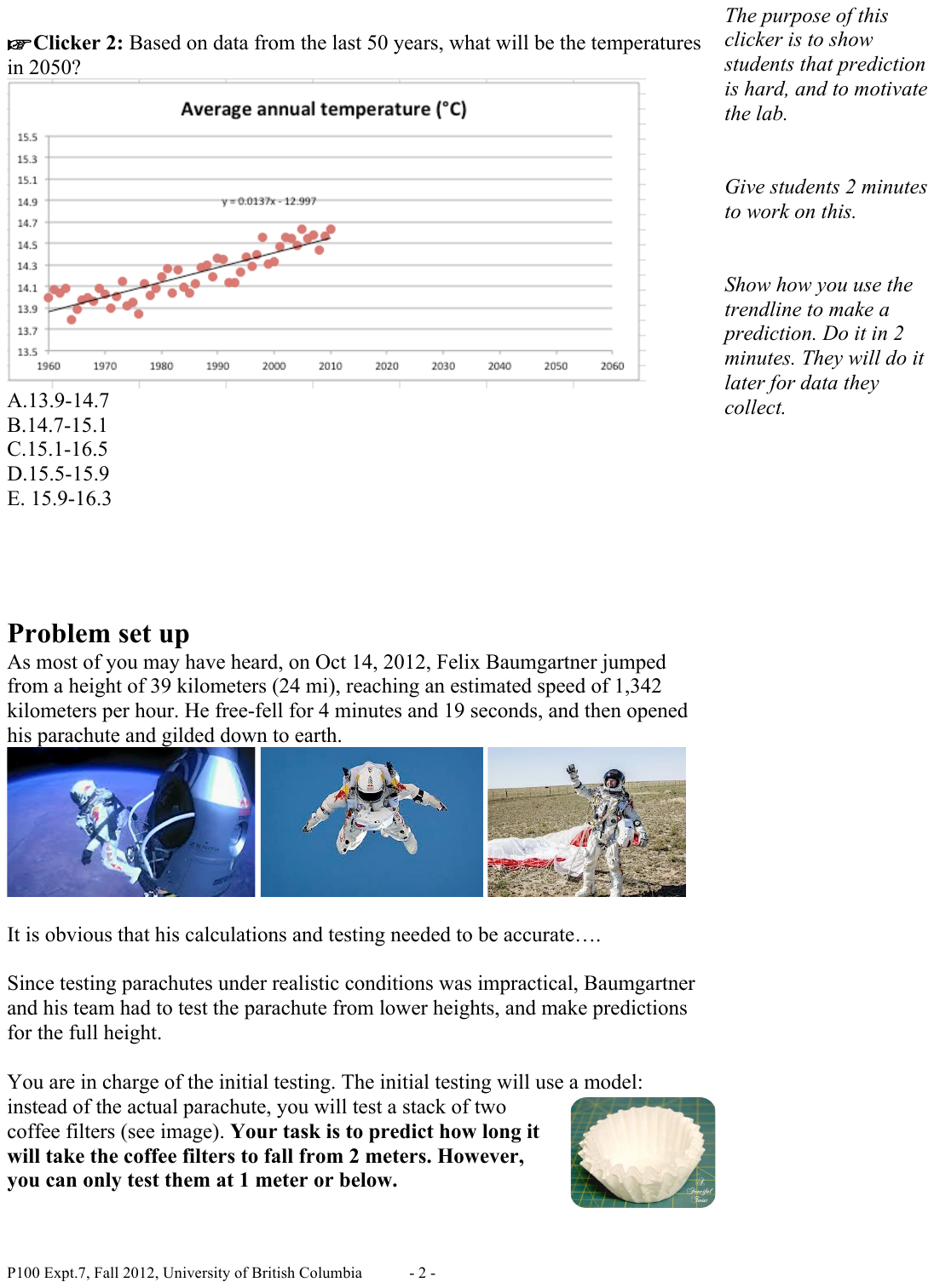}}
\caption{}
\label{}
\end{figure*}

\newpage
\begin{figure*}[h!]
\centering
\fbox{\includegraphics[scale=0.65]{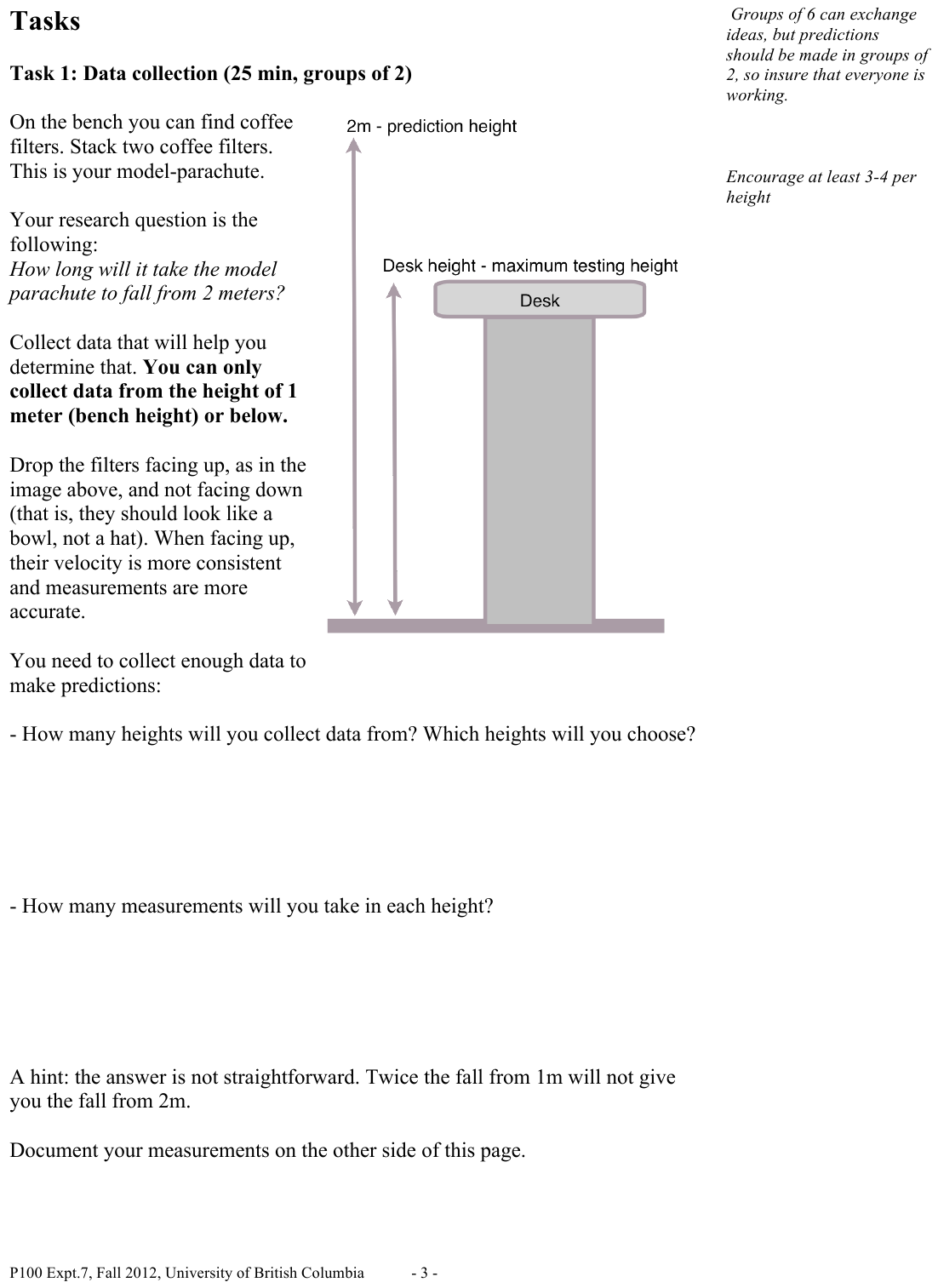}}
\caption{}
\label{}
\end{figure*}

\newpage
\section{TA observation form}
\label{TA_form}

\begin{figure*}[h!]
\centering
\fbox{\includegraphics[scale=0.65]{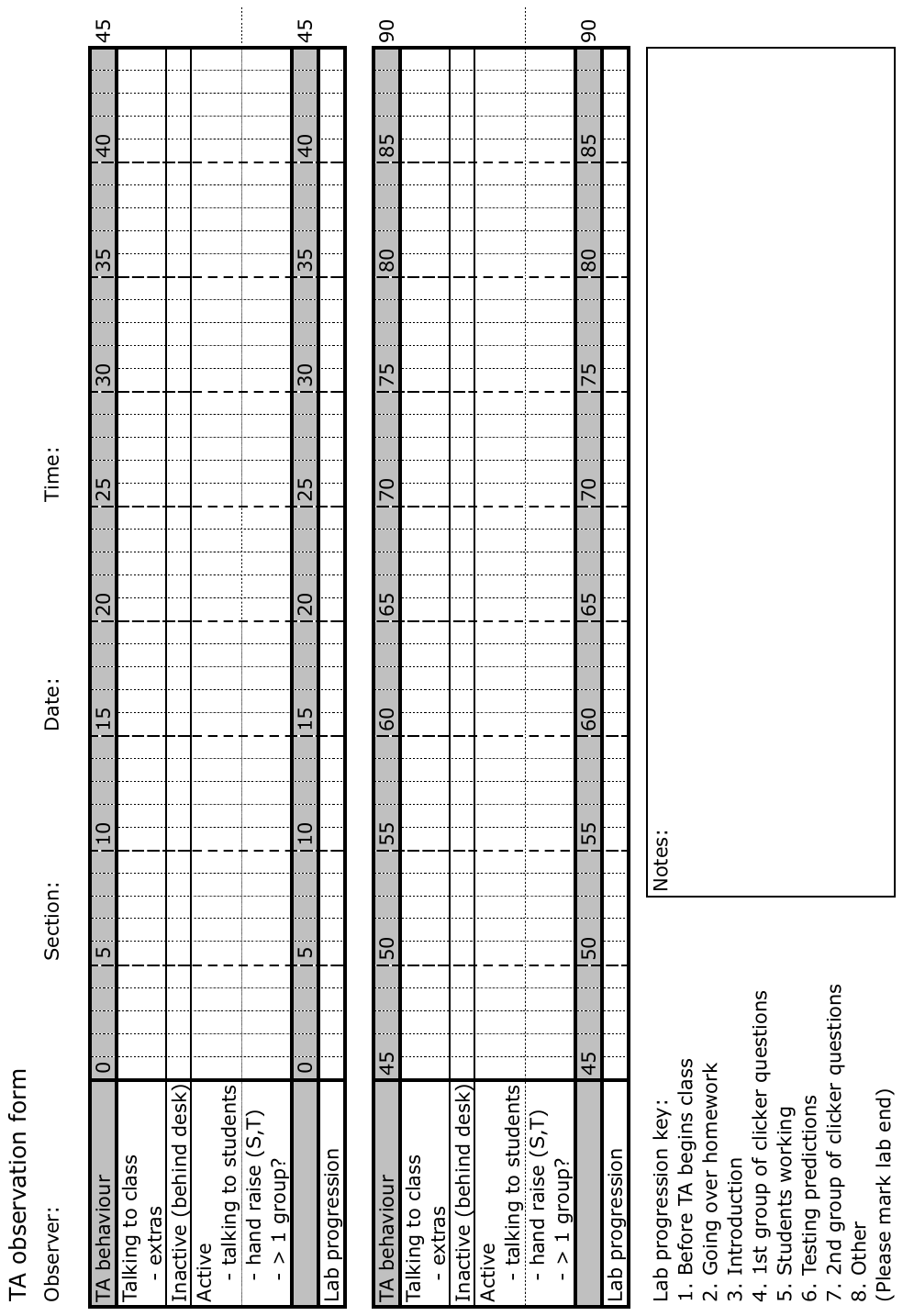}}
\caption{The TA observation form. To complete the form, the observer shades in the areas of the timeline corresponding to the TA's actions during the lab, giving a record of the TA's behaviors during the lab.}
\label{TAform}
\end{figure*}

\newpage
\section{Recorded TA behaviors that were outside the script of the lab}
\label{extras}
Here, we give a list of the distinct TA behaviors that were noted as `extras' by the observers.
\begin{itemize}
\item Interrupted class to make announcement.
\item Gathered class at the front chalkboard.
\item Pointed out error in handout by using overhead.
\item Going over method of solving clicker.
\item Asked other TA question in front of everyone to spark explanation.
\item Using chalkboard to explain previous week's homework.
\item Asking questions to students during the explanation.
\item Asking students if material is clear.
\item Students discussing clicker questions with peers.
\item Soliciting student responses in a class discussion.
\item Using chalkboard for an explanation in an individual interaction.
\item Explaining clicker question at chalkboard.
\item Working on overhead to show predicted versus average.
\item Other TA comments during class discussion.
\item TA jumps in while other TA is leading a discussion.
\item Showed video clip to motivate lab.
\item Banter with TA partner during class discussion.
\item TA says to class: `Today is my favorite day. I hope it's yours too.'
\item Impromptu discussion with the chalkboard about some student questions.
\item Playing music during the lab.
\item Used projector.
\item Used Matlab.
\item Brief Excel tutorial on computer.
\item Gave students 30 seconds to individually discuss a question during a classroom discussion.
\item Using in-class cameras.
\item Referring to material shown on projector.

\end{itemize}

\newpage
\section{On/off task form}
\label{oo_form}

\begin{figure*}[h!]
\centering
\fbox{\includegraphics[scale=0.65]{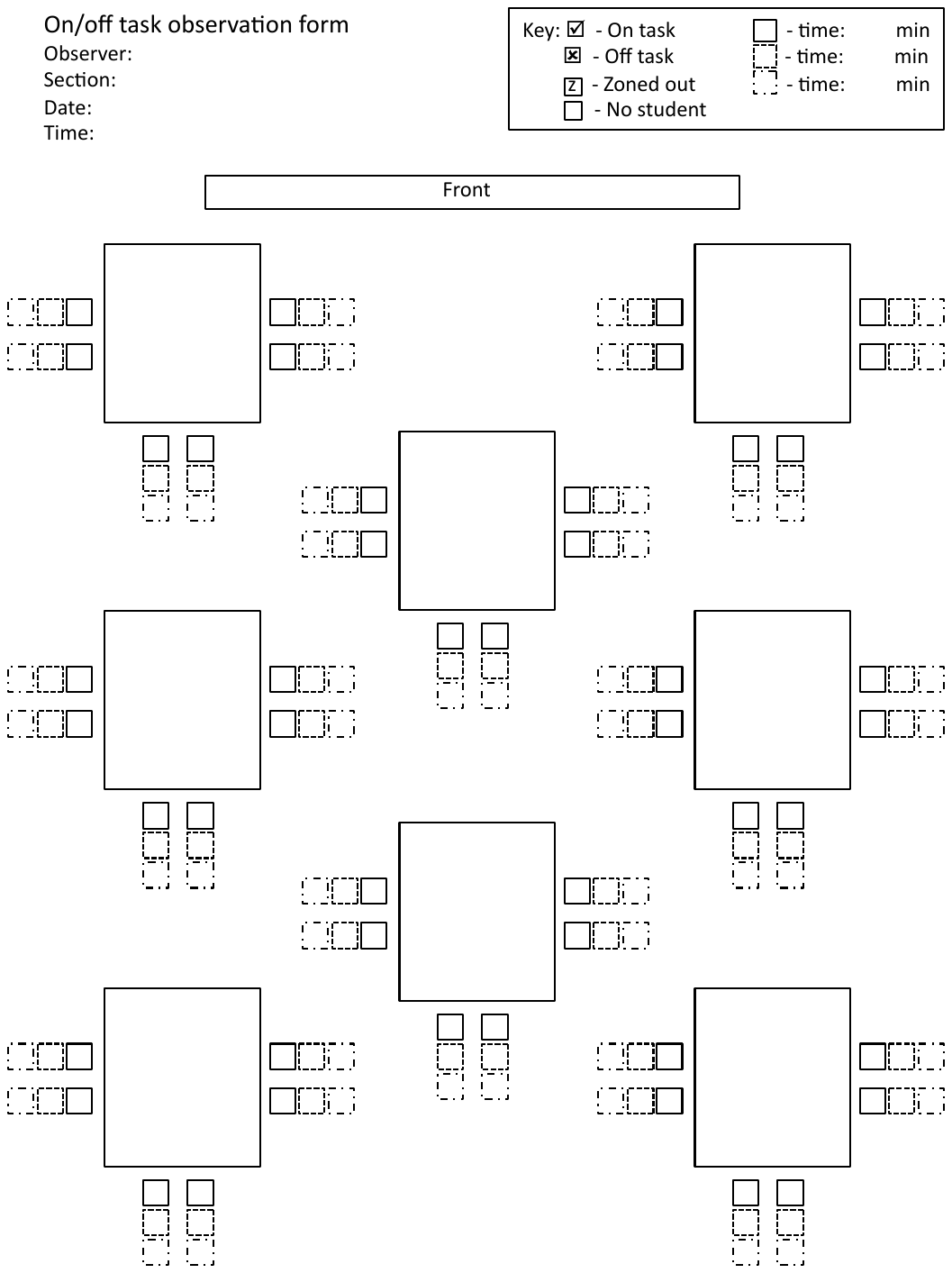}}
\caption{The On/off task form. The form is a spatial map of the lab on which the observer records student engagement. Each form allows three snapshots of engagement to be recorded.}
\label{TAform}
\end{figure*}

\section{Example questions from the lab exam}
\label{lab_test}
These are three example questions from the lab skills exam given during the first and last week of the lab.

\begin{enumerate}
\item Three environmentalists want to evaluate whether summers in Vancouver got warmer during the 20th century (1900-2000). They can choose one of the following data sets. Which data set should they analyze?
\begin{enumerate}
\item 1980, 1985, 1990, 1995, 2000
\item 1906, 1907, 1908, 1909, 1999
\item 1900, 2000
\item 1920, 1940, 1960, 1980
\end{enumerate}

\item John and Lesley measured the length of the corridor in their dorms. Each of them measured the distance three times: John measured: 10 m, 85 m, 43 m. Lesley measured: 43 m, 45 m, 44 m. Which of the following values is the closest to the actual length of the corridor?
\begin{enumerate}
\item 43 m, the only result that repeats more than once.
\item 44 m, the average of Lesley.
\item 45 m, the average of all values
\item 46 m, the average of John
\item 47.5 m, the middle between the lowest (10) and the highest (85)
\end{enumerate}

\item Dave and Jill measured the friction coefficient between two blocks of the same material. According to the textbook, the coefficient for two pieces of wood is 0.4. They argued how many times they should measure the coefficient until they can stop measuring. Which of the following answers is most correct?
\begin{enumerate}
\item After they receive the same values twice.
\item When they notice that results converge to a single range.
\item After two measurements.
\item When they receive 0.4.
\end{enumerate}

\end{enumerate}

\end{appendix}

\begin{acknowledgments}

We gratefully acknowledge Doug Bonn and Georg Rieger for useful discussions and Natasha Holmes and Firas Moosvi for comments on the manuscript. This work was funded by the Carl Wieman Science Education Initiative.

\end{acknowledgments}

\bibliography{TA_refs}

\end{document}